\documentclass[apj]{emulateapj}
\pdfoutput=1
\usepackage{amsmath}
\usepackage{ulem}
\usepackage{CJK}
\usepackage{epsfig}
\usepackage{epstopdf}

\shorttitle{Polarization of magnetized reverse shock emission} \shortauthors{Deng et al.} \slugcomment{}
\begin{document}
\begin{CJK*}{UTF8}{gbsn}
\title{
Magnetized reverse shock: density-fluctuation-induced field distortion, polarization degree reduction, and application to GRBs}
\author{Wei Deng (邓巍)\altaffilmark{1,2}, Bing Zhang (张冰)\altaffilmark{1}, Hui Li (李晖)\altaffilmark{2}, James M. Stone\altaffilmark{3}}

\affil{\altaffilmark{1}Department of Physics and Astronomy, University of Nevada Las Vegas, Las Vegas, NV 89154, USA; deng@physics.unlv.edu, zhang@physics.unlv.edu}
\affil{\altaffilmark{2}Los Alamos National Laboratory, Los Alamos, NM 87545, USA; hli@lanl.gov}
\affil{\altaffilmark{3}Department of Astrophysical Sciences, Princeton University, Princeton, NJ 08544-1001, USA; jstone@astro.princeton.edu}

\begin{abstract}
The early optical afterglow emission of several gamma-ray bursts (GRBs) shows a high linear polarization degree (PD) of tens of percent, suggesting an ordered magnetic field in the emission region. The light curves are consistent with being of a reverse shock (RS) origin. However, the magnetization parameter, $\sigma$, of the outflow is unknown. If $\sigma$ is too small, an ordered field in the RS may be quickly randomized due to turbulence driven by various perturbations so that the PD may not be as high as observed.
Here we use the ``Athena++" relativistic MHD code to simulate a relativistic jet with an ordered magnetic field propagating into a clumpy ambient medium, with a focus on how density fluctuations may distort the ordered magnetic field and reduce PD in the RS emission for different $\sigma$ values. For a given density fluctuation, we discover a clear power-law relationship between the relative PD reduction and the $\sigma$ value of the outflow. Such a relation may be applied to estimate $\sigma$ of the GRB outflows using the polarization data of early afterglows.
\end{abstract} 

\keywords{gamma-ray burst: general - galaxies: jets - polarization - magnetic fields - shock waves - magnetohydrodynamics (MHD)}

\section{Introduction\label{sec:intro}}

The jet composition and energy dissipation, particle acceleration and radiation mechanisms of GRBs have not been identified and thus are subject to debate \citep[][for a recent review]{Kumar&Zhang15}. Leading models include the matter-dominated fireball shock models \citep[e.g.][]{Goodman86,Paczynski86,Rees94,Meszarosrees00} and the Poynting-flux-dominated magnetic dissipation models \citep[e.g.][]{Lyutikov03,Zhang&Yan11,Mckinney&Uzdensky12,Deng15,Deng16}. Growing evidence suggests that the GRB jets carry a dynamically important magnetic field, so that magnetic reconnection may play an important role in producing observed GRB prompt emission and afterglow emission.

One piece of such evidence comes from the observations of the early optical afterglow of GRBs, which is dominated by a rapidly decaying ($\propto t^{-2}$) segment early on from the external reverse shock (RS) \citep{Akerlof99,Meszaros&Rees99,Sari&Piran99,Gomboc08,Gao15}. Modeling such a component suggests that the RS is more magnetized than the forward shock (FS), so that the GRB jet is more magnetized than the ambient medium \citep{Fan02,Kumar03,Zhang03,Zhang05}. This is consistent with the scenario that prompt emission is powered by magnetic reconnection and turbulence with a moderate magnetization parameter $\sigma$ in the prompt emission region \citep{Zhang&Yan11}. On the other hand, when the jet is decelerated by an ambient medium (when the RS propagates through the ejecta), the $\sigma$ value of the ejecta cannot be very high. Otherwise, the RS would be weak or suppressed \citep{Zhang05,Mimica09,Mimica10,Mizuno09}, contrary to the observations.

Direct proof of such a scenario comes from the polarization observations of the early optical emission of some GRBs. \cite{Steele09} discovered an $\sim 10\%$ polarization degree (PD) of the early RS optical emission from GRB 090102. \cite{Mundell13} for the first time measured the ``polarization light curve'' of GRB 120308A,  which shows an $\sim 28\%$ PD at 4 minutes after the GRB, which is reduced to $\sim 16\%$ PD over the subsequent 10 minutes as the FS emission contribution becomes stronger. These observations suggest that an ordered magnetic field in the RS region exists \citep{Lan16}. However, the $\sigma$ value of the RS emission cannot be easily diagnosed based on the PD and light-curve data.

For a relativistic outflow, strong turbulence is likely developed in the flow due to various perturbation mechanisms (e.g. density clumps in the ambient medium or intrinsic irregularity within the flow). If $\sigma \ll 1$, even if an ordered magnetic field configuration may exist initially, turbulence may quickly randomize the field configurations, which would greatly reduce the ordered field component and the PD of synchrotron radiation. As a result, based on the observed PDs of GRBs 090102 and 120308A, one may be able to place a {\em lower limit} on the $\sigma$ values of the outflows. To do this, detailed numerical simulations are needed. 

The main goal of this work is to investigate the resilience of the ordered magnetic fields in a relativistic jet against the density-fluctuation perturbations from the ambient medium as well as the resultant reduction of the PD of synchrotron radiation due to such perturbations. To achieve this goal, we have chosen to solve a simplified problem, i.e. to simulate how an ambient density fluctuation alters the initially parallel magnetic fields in a relativistic jet. A more realistic problem would invoke a partially ordered magnetic field configuration (after the GRB prompt emission phase, which resulted from dissipation of magnetic fields) in a conical jet (in which a Rayleigh-Taylor (RT) instability would play a role to further destroy ordered magnetic fields). In any case, our results present 
 %based on a simplified but relatively stable configuration\footnote{\bf E.g. the purely parallel initial magnetic field, no RT instability, only including the ambient density fluctuation. (See the detailed description in Section 2.)} from the turbulence generation and the PD reduction point of view, compared with most of the possible complicated realistic situations. This will give 
a ``lower limit" of the relative PD reduction for a certain $\sigma$ value, which may be potentially used to constrain the ``lower limit" of $\sigma$ in a GRB outflow once a PD is measured. 
% based on future observations of the relative PD change between the end of the prompt emission phase and the early stage of the RS dominant early afterglow phase. 
The simulations are presented in Section 2. The simulation results, especially a relationship between the relative PD reduction and the initial $\sigma$ value of the ejecta are presented in Section 3. The results are summarized in Section 4 along with a discussion on their applications to the GRB problem.
%, we discuss how to use our results to constrain the $\sigma$ value of the ejecta.

\section{Problem setup\label{sec:setup}}

We use the relativistic MHD code ``Athena++" \citep{White16} to simulate the dynamical evolution of the outflow, and the ``3DPol'' multi-zone polarization-dependent ray-tracing radiation code \citep{Zhang14} to model the PD of synchrotron radiation in the RS region.

We perform our simulations in the ``lab'' frame. We set up a jet that propagates in the $z$ direction with a relativistic speed into an ambient medium at rest (Fig. 1). In order to isolate the problem in testing the resilience of ordered magnetic fields against density perturbation, we adopt several simplified assumptions. First, the jet has a planar geometry, i.e., the cross-section of the jet does not expand with time (as is the case of a conical jet). This would reduce other factors, such as the RT instability \citep{Dufell&MacFadyen13}, to deform ordered magnetic fields. Second, we assume a parallel magnetic field configuration within the simulation domain in the unshocked region. In principle, after prompt emission, the magnetic field configuration is expected to be somewhat distorted. Adopting a simple magnetic field configuration, on the other hand, can remove other effects and allow us to concentrate on the density perturbation effect on destroying the ordered fields. The global magnetic field configuration of a GRB ejecta may not be approximated as the simple parallel configuration we use. However, a GRB observer only observes emission from a $1/\Gamma$ cone, in which a toroidal magnetic field configuration \citep[e.g.][]{Lyutikov03b,Deng15} can be approximated as being approximately parallel \citep[e.g.][]{Lazzati06}.

For the planar geometry problem, it is convenient to simulate the jet propagation problem using a 2D rather than a 3D box\footnote{Even though the simulation box is essentially 2D, all the vectors are solved in 3D.}. The effective resolution of the simulations is $\Delta y \approx \Delta z \approx 0.0012 L_0$, where $L_0$ is the length normalization factor between code units and physical units. The entire simulation frame has $20 L_0$ in the $z$ direction and $0.156 L_0$ in the $y$ direction. Since the simulation box may be regarded as a patch inside the $1/\Gamma$ cone of the entire thin shell in both of the radial and the transverse directions, we set up a periodic boundary condition in the $y$ direction and an inflow boundary condition in the lower boundary of the $z$ direction. 

As the jet runs into the medium, four regions are developed: (1) unshocked medium, (2) shocked medium, (3) shocked jet, and (4) unshocked jet. Our initial conditions define the parameters in regions 1 and 4. We use the subscripts ``4" and ``1" in the following parameters to denote the unshocked jet and the unshocked medium, respectively, $\Gamma_4$ is the initial jet Lorentz factor, $n$ is the number density, and $P$ is the thermal pressure. The parameters are all in the code units (Table \ref{tab:para}). One can convert them to the physical units by assigning $L_0$ and  the typical magnetic field strength $B_0$ to certain values.  For example, if one takes a typical physical parameters $L_0 = 3\times 10^{13}$ cm, one then has the time normalization parameter $t_0=L_0/c = 10^{3}$ s in the frame, which corresponds to $t_{\rm obs} \sim t_0 / \Gamma_4^2 = 10$ s in the observer frame. Our simulation lasts for $19 t_0$, which corresponds to $\sim 190$ s in the observer frame, a time scale shorter than the deceleration time (or RS crossing time) for $\Gamma_4=10$. This is also consistent with the estimate that the simulation scale $20 L_0 = 6 \times 10^{14}$ cm is shorter than the thickness of the shell $R_{\rm dec}/\Gamma^2 = (10^{15} ~{\rm cm}) R_{\rm dec,17}  \Gamma_{4,1}^{-2}$, so that the simulation box is only part of the global jet.

%\footnote{\bf We consider that the thickness of the thin shell is roughly around ($10^{16 - 17}$ cm)/$10^2$ $\sim 10^{14-15}$ cm in the simulation frame (lab frame) with the Lorentz factor $\Gamma=10$ (introduced later), which should be larger than $20 L_0$ (box length), so $L_0 \lesssim 5\times 10^{12-13}$ cm.}.

For a conical jet at the RS crossing time (which is also the deceleration time), the shock jump conditions at both FS and RS gives a relation \citep{Sari&Piran95,Zhang05}\footnote{To derive this formula, the adiabatic index $\hat\gamma$ has been assumed to be $\Gamma$-dependent, i.e. $\hat\gamma = (4\Gamma+1)/3\Gamma$ \citep[e.g.][]{Kumar&Granot03,Uhm11}.}
\begin{equation}
 \frac{n_4}{n_1} =\frac{\Gamma_{21}^2-1}{\Gamma_{34}^2-1},
\end{equation}
where $\Gamma_{21}$ and $\Gamma_{34}$ are the relative Lorentz factors between upstream and downstream at the FS and RS, respectively. For a thin shell ($\Gamma_{34} \gtrsim 1$), which is relevant for most GRB problems, one gets a rough relation $n_4/n_3 \sim \Gamma_4^2$. Since our simulation does not introduce a conical jet (for which $n_4/n_1$ decreases with time), we set up $n_4/n_1 = \Gamma_4^2$ by hand to mimic the physical condition at the deceleration time.

The initial setup of the number density distribution
%\footnote{\bf We use the same setup of the density and pressure profiles as \cite{Mizuno09} did, who consider $n_4/n_1 = [(\Gamma_2-1)(4\Gamma_2+3)]/[(\Gamma_{34}-1)(4\Gamma_{34}+3)] \sim \Gamma_4^2$, empirically \citep{Sari&Piran95,Zhang05}, where $\Gamma_{34}$ is the relative Lorentz factor between the shocked and the un-shocked jet.} 
is shown in Fig. 1a. The high density ejecta ($n_4=100$) and the low density ambient ($n_1=1$) are separated at the position $z=1$. Some density clumps are introduced in the ambient medium to mimic density fluctuation. The density profile of each clump is delineated in the form $n = n_1+C \cdot e^{-r}$, where r is the distance away from the center of the clump, and the displacement between the centers of two neighboring clumps are $y_{disp}=0.04L_0$ and $z_{disp}=2L_0$, and $C=100$ is the amplitude of the fluctuation\footnote{The degree of distortion depends on the distribution and amplitude of the clumps. A denser distribution of the clumps tends to distort the RS more significantly, leading to a larger PD reduction for a certain initial $\sigma$ value.  We choose a relatively sparse distribution of the clumps to set a more conservative ``lower limit'' of the relative PD reduction.}.
The initial magnetic field configuration is assumed to be uniform in the jet, with $B_{y,4}$ only in the co-moving frame. In the lab frame, an electric field component in the $x$ direction ($E_{x,4}$) with a similar amplitude as $B_{y,4}$ (self-consistently determined by $B_{y,4}$ and $\Gamma_4$) is introduced so that a Poynting flux is coupled with the jet in the lab frame. 
%As we have discussed above, since the global structure of the jet does not matter much for an observer within the $1/\Gamma$ cone, we introduce a uniform magnetic field ($B_{y,4}$ only) perpendicular to the direction of motion in the jet (Fig. 1c)\footnote{\bf In the co-moving frame, the ejecta carries a uniform $B'_{y,4}$. In the lab frame, one sees a Poynting-flux, with an electric field component in the $x$ direction ($E_{x,4}$) with a similar amplitude as $B_{y,4}$. By setting $B_{y,4}$ and $\Gamma_4$ in the lab frame, the $E_{x,4}$ component is self-consistently incorporated.}. 
Different values of $B_{y,4}$ are explored, which correspond to different initial values of the magnetization factor $\sigma_4$ in the ejecta. Since the initial thermal energy is much smaller than the kinetic energy, $\sigma_4$ can be calculated as $\sigma_4 \approx B_{y,4}^2/4 \pi \Gamma_4^2 n m_p c^2$, where $m_p$ is the proton mass and $c$ is the speed of light.
The typical Lorentz factor ($\Gamma_4$) of GRBs at the deceleration phase is around several hundred (e.g. \citealt{Liang10}). Such a large Lorentz factor is not easy to achieve for stable simulations. We set up two sets of simulations with $\Gamma_4 = 10, 20$, respectively\footnote{The maximum $\Gamma$ that the code can handle is higher and varies depending on the values of other parameters, e.g. $\sigma$.} (Fig. 1b). Although these values are smaller than the true Lorentz factor of GRBs at the deceleration phase, the simulation can nonetheless catch the key physics explored in this paper. The two different values $\Gamma_4$ adopted in our simulations can reflect the dependence of results on the initial Lorentz factor.

The parameters for all the simulation models are listed in Table 1. Model 0 is the reference model without density fluctuation and magnetic field in the jet, which is used to test the basic evolution process of the blastwave. The simulation results are fully consistent with analytical solutions, suggesting the validity of relativistic Athena++. Models 1-5 have the same density fluctuations but different $\sigma_4$ (and hence $B_{y,4}$) in the ejecta. This allows us to explore how magnetic field distortion and the resulting PD vary with $\sigma_4$. Models 6-10 have a different initial Lorentz factor ($\Gamma_4$=20) corresponding to Models 1-5 ($\Gamma_4$=10), which is a preliminary test of the effect of $\Gamma_4$.

The $\sigma_4$ values are set to be below unity based on two considerations. (1) Significant magnetic dissipation has likely occurred during the prompt emission phase \citep{Zhang&Yan11,Deng15}, so that the $\sigma$ of the outflow has significantly reduced beyond the prompt emission radius. There is a continuous decrease of $\sigma$ as the jet expands and at the deceleration radius, $\sigma_4$ is likely below unity \citep[e.g.][]{GaoZhang15}. (2) The RS is weak or does not exist if $\sigma_4$ is above unity \citep{Zhang05,Mimica09,Mizuno09}. Since observationally one does observe the RS emission, $\sigma_4$ cannot be too high.

\begin{figure}[!htb]
%\plotone{fig1.pdf}
\begin{center}
\includegraphics[angle=0,scale=0.31]{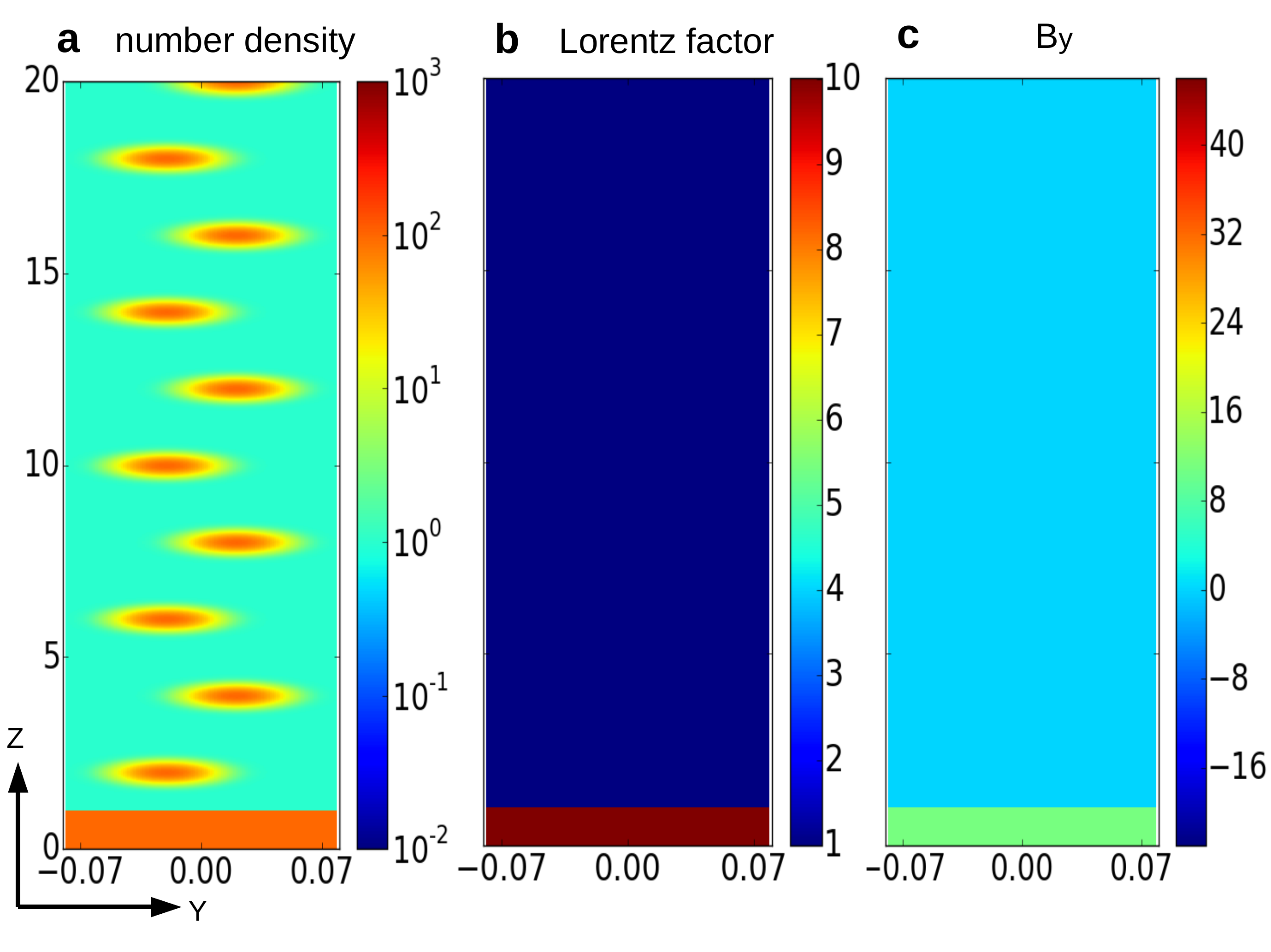}
\caption{Initial setup of the number density distribution ({\bf a}), the Lorentz factor distribution ({\bf b}), and the initial magnetic field $B_y$ distribution ({\bf c}) for Model 3. The front of the jet is located at $z=1$ initially and propagates upwards to the ambient medium.}
\label{fig:ini}
\end{center}      
\end{figure}

\begin{table}
\scriptsize 
%\begin{table}[!htb]
\centering
\caption{The initial parameters of the simulation models}
\begin{tabular}{|c||c|c|c|c|c|c|c|}
\hline
\multicolumn{8}{|c|}{Group 1: $\Gamma_4=10$}\\\hline
Models & $\sigma_4$ & $B_{y,4}$ & $\Gamma_4$ & $n_4$ & $n_1$ & $P_4$ & $P_1$ \\\hline
0 & 0 & 0 & 10 & 100 & 1 & 1 & 0.01\\
1 & 0.1 & 31.62 & 10 & 100 & 1 & 1 & 0.01\\
2 & $10^{-2}$ & 10 & 10 & 100 & 1 & 1 & 0.01\\
3 & $10^{-3}$ & 3.162 & 10 & 100 & 1 & 1 & 0.01\\
4 & $10^{-4}$ & 1 & 10 & 100 & 1 & 1 & 0.01\\
5 & $10^{-5}$ & 0.3162 & 10 & 100 & 1 & 1 & 0.01\\
\hline
\multicolumn{8}{|c|}{Group 2: $\Gamma_4=20$}\\\hline
Models & $\sigma_4$ & $B_{y,4}$ & $\Gamma_4$ & $n_4$ & $n_1$ & $P_4$ & $P_1$ \\\hline
6 & 0.1 & 63.24 & 20 & 100 & 1 & 1 & 0.01\\
7 & $10^{-2}$ & 20 & 20 & 100 & 1 & 1 & 0.01\\
8 & $10^{-3}$ & 6.324 & 20 & 100 & 1 & 1 & 0.01\\
9 & $10^{-4}$ & 2 & 20 & 100 & 1 & 1 & 0.01\\
10 & $10^{-5}$ & 0.6324 & 20 & 100 & 1 & 1 & 0.01\\
\hline
\end{tabular}\\
\label{tab:para}
\end{table}

\section{results\label{sec:results}}

\subsection{The representative case: Model 3\label{subsec:case3}}

We first present the detailed evolution of the representative case of Model 3 ($\sigma_4=10^{-3}$). The zoom-in contour plots of the number density (Fig. 2b) and the two components of the magnetic field, $B_y$ (Fig. 2c) and $B_z$ (Fig. 2d) are presented at $t=8 t_0$, where $t_0 = L_0/c$ is the time normalization factor between the code and physical units. 
As a comparison, we also show the density contour of Model 0 in Fig. 2a. One can see that the surface of the contact discontinuity (CD, the separation between the yellow and the red regions) is highly distorted in Model 3 due to the perturbation of the density clumps. Also, in the RS region, the initially pure ordered magnetic field $B_y$ is significantly distorted and a significant $B_z$ component is generated. This significant topological change of the magnetic field would inevitably lead to the reduction of the synchrotron radiation PD in the RS emission.

\begin{figure}[!htb]
%\plotone{fig2.pdf}
\begin{center}
\includegraphics[angle=0,scale=0.3]{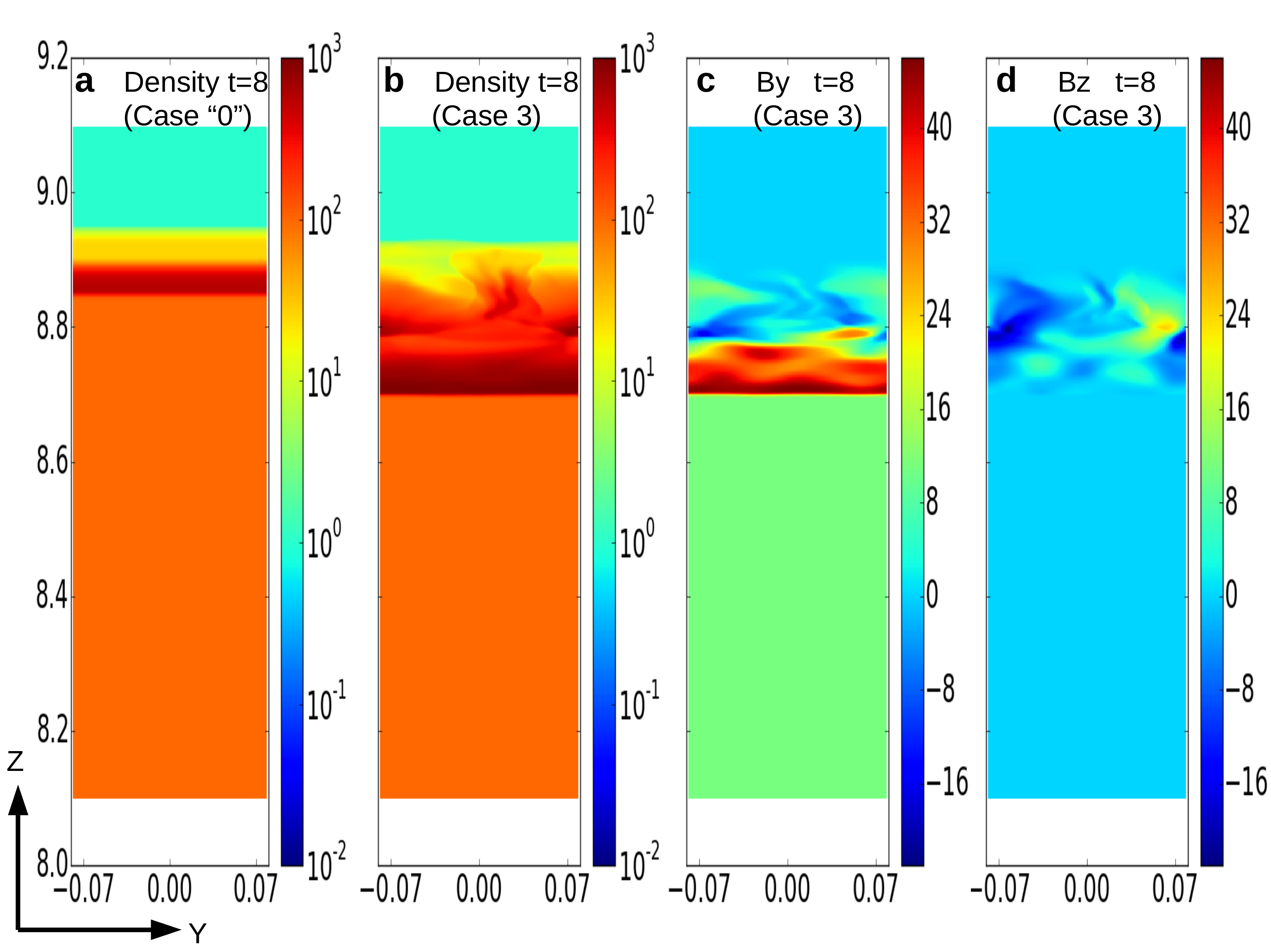}
\caption{Zoom-in cuts of our simulation results for Model 3 ($\sigma_4=10^{-3}$) at t=8 $t_0$. Panel ({\bf a}) is the density contour plot of Model 0 as a reference. Panels ({\bf b}), ({\bf c}), and ({\bf d}) are the number density, $B_y$, and $B_z$ contour plots of Model 3. The magnetic field lines are significantly distorted by the density fluctuations, and a significant $B_z$ component is developed.}
\label{fig:case3_cut}
\end{center}      
\end{figure}

Figure 3 shows the synchrotron radiation PD evolution of Model 3. To calculate the polarization evolution, we use the same method used in our previous work \citep{Deng16,Zhang16}. We first define the region between the CD and the RS in the surface density contour as the region where RS emission comes from. In order to handle the highly distorted CD surface, we develop a numerical approach to define the complex CD profile based on a sharp density discontinuity in the numerical cells. We then uniformly inject particles with a non-evolving power-law energy distribution ($N_e =  N_{e,0} \gamma_e^{-2.5}$, where $\gamma_e \in (10, 10^6)$ and $N_{e,0}$ is an arbitrary normalization factor\footnote{We do not simulate the light curve and spectral evolution in our radiation calculation. The simulated PD and PA evolutions do not depend on the particle number normalization factor ($N_{e,0}$), which can be taken as an arbitrary number in our polarization calculations.}) in the entire RS region. Next, we use the ``3DPol" multi-zone polarization-dependent ray-tracing radiation code \citep{Zhang14} to calculate the evolution of the PD from the RS region based on our MHD simulation results. The polarization calculations are carried out by post-processing the MHD simulation results with particles injected every two time steps in the MHD simulation.
From Fig. 3, we show that the PD quickly (in $\sim 2 t_0$) drops from the initial high value to a lower value. This is caused by the distortion of magnetic fields due to density fluctuations. In long term, the PD gradually recovers due to the resilience of the magnetic field lines. However, we believe that such a recovering effect may be artificial. By considering a more realistic global magnetic configuration \citep{Deng15} and the Rayleigh-Taylor instability in the conical shell  \citep{Dufell&MacFadyen13}, one would expect that the distorted $B$ configuration may not be easily recovered. In the following discussion, we focus on the minimum value in the PD curve and use it to define the reduction of PD from the original value.

\begin{figure}[!htb]
\begin{center}
\includegraphics[angle=0,scale=0.6]{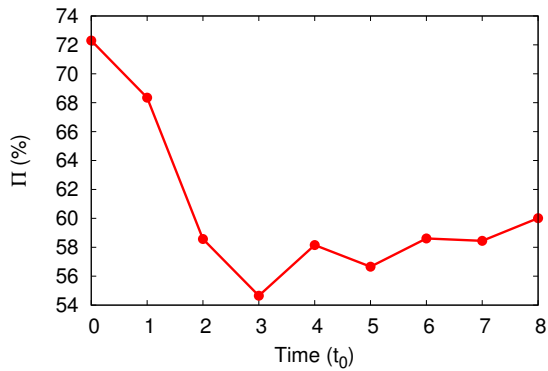}
\caption{Polarization evolution of Model 3 ($\sigma_4=10^{-3}$). It shows a significant PD drop initially due to the distortion of the magnetic field by the ambient medium density fluctuation. The later PD gradually increases, but it is mainly due to the simplification of our model and most likely does not exist in the real situations.}
\label{fig:case3_pol}
\end{center}      
\end{figure}

\subsection{Relationship between PD reduction and $\sigma$ \label{subsec:relationship}}

We run other Models with different initial $\sigma_4$ for each group of $\Gamma_4$, and display the initial PD and final PD in Fig. 4a. For the uniform magnetic field distribution and a planar geometry, the maximum PD allowed for synchrotron radiation, i.e. $\Pi=(p+1)/(p+7/3) \simeq 72\%$ for $p=2.5$ \citep{Rybicki79}, can be achieved. After density perturbation, one can see that the minimum PD in the RS region clearly scales with $\sigma_4$ for each group of $\Gamma_4$. The smaller the $\sigma_4$ value, the lower the minimum PD. In reality, the field configuration after the prompt emission phase must have been distorted due to magnetic dissipation processes such as internal-collision-induced magnetic reconnection and turbulence \citep{Zhang&Yan11,Deng15}, the initial $\Pi$ in the RS crossing phase must be lower than 72\%. As a result, the {\em relative PD reduction}, defined as $\xi = (\Pi_0 - \Pi_{\rm min})/\Pi_0$, may be more relevant. This is plotted in Fig. 4b, which clearly shows a decrease of the relative reduction with increasing $\sigma_4$ for each group of $\Gamma_4$. Fits to this power-law regime ($\sigma_4=0.1$ removed) give empirical relations $\xi=6.0 \sigma_4^{-0.21}$ for the group of $\Gamma_4=10$, and $\xi=1.0 \sigma_4^{-0.37}$ for the group of $\Gamma_4=20$.

The behavior may be understood in terms of pressure and energy. One may consider the problem in the rest-frame of the RS region, in which the RS region is a target with growing thickness (as the RS front propagates into the jet) while the density clumps are bullets penetrating into the target. In our simulations, the ram pressure ($P_r$) and kinetic energy (${\cal E}_K$) of the density clumps remain the same for different models. The co-moving magnetic pressure ($P_B = {B'}^2/8\pi$) varies with $\sigma$ in different models, and the co-moving magnetic energy (${\cal E}_B$) varies with $\sigma$ and also with time as the volume of the RS region expands. In all our models, $P_r$ is designated to be greater than $P_B$, so that the magnetic field configuration can be distorted.
When $\sigma$ is small, the magnetic field is dynamically unimportant. The magnetic amplification factor across the RS front depends on $\Gamma$ only when the initial density and pressure ratio are fixed, so that the magnetic pressure in the RS region would scale linearly with $\sigma$ in logarithmic space for each group of $\Gamma_4$. This is indeed seen from the numerical data in Fig. 4c.
Since the magnetic pressure is the main stabilizer of the magnetic configuration against density-fluctuation perturbation, and the relative PD reduction therefore follows a power-law relation with $\sigma$ in the $\sigma \ll 1$ regime.
We also find that the power-law indices are different between two groups with the different initial Lorentz factor, which means that the indices may be a function of the initial Lorentz factor, the initial density and pressure ratio, and the amplitude of the ambient density perturbation, which need additional investigation in the future.

\begin{figure}[!htb]
\begin{center}
\includegraphics[angle=0,scale=0.6]{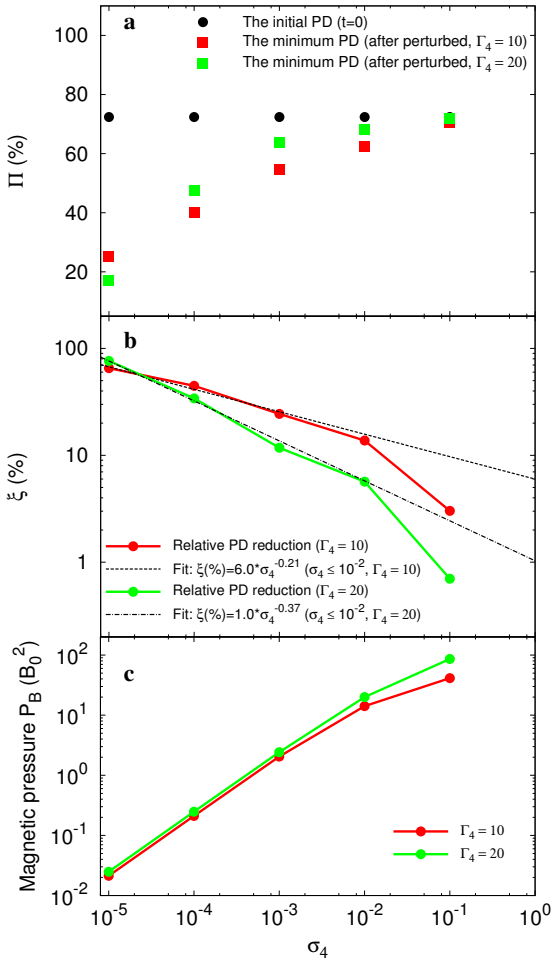}
\caption{({\bf a}) The initial and the minimum values of PD during the PD evolution for different initial $\sigma_4$, ({\bf b}) the relationship between the relative PD reduction and initial $\sigma_4$, and ({\bf c}) the relationship between the magnetic pressure in the co-moving frame of the RS and $\sigma_4$. Both show a power-law relationship when $\sigma_4$ is relatively small ($\sigma_4<0.01$).}
\label{fig:pol_sigma}
\end{center}      
\end{figure}

When $\sigma_4$ is higher than 0.01, the magnetic field starts to affect the RS dynamics, which would decrease the RS strength and the compression ratio and in the meantime increase the reverse shock speed in the CD frame. Although the magnetic pressure increases more slowly (Fig. 4b) due to the weaker RS, the thickness of the RS region and hence the magnetic energy increase more rapidly due to the faster RS speed \citep{Zhang05,Mimica09,Mizuno09}. Even though the field configuration close to CD is distorted more, the field configuration from the far end of CD (close to the RS front) is less perturbed due to the larger volume of the RS region so that, on average, the field configuration in the entire RS region is less perturbed. This explains the roll-over in the $P_B-\sigma$ relation in Fig. 4c, and the relatively small PD reduction at higher $\sigma$ in Fig. 4b.

\section{conclusions and discussion\label{sec:conclusion}}

In this paper, we explore how resilient a magnetic field is in a relativistic jet against density perturbation from the ambient medium in view of the recent observations of moderately high linear PD in some GRB early optical afterglows \citep{Steele09,Mundell13}. The motivation is to set a possible lower limit on the $\sigma$ value of the outflow based on the data. In view of the complication of the GRB RS problem, we limit ourselves to a cylindrical-like jet with a planar geometry and an initially uniform ordered magnetic field configuration. By introducing a perturbation due to density fluctuations from the ambient medium, we investigate how the distortion of magnetic configuration and reduction of PD depend on the $\sigma$ parameter of the jet. As expected, our numerical simulations using the relativistic Athena++ code show more significant reduction of PD with decreasing $\sigma$. The numerical results show an interesting power-law behavior between the relative PD reduction and $\sigma$, which may be applied to place constraints on the jet composition of GRBs. 
For example, if at the end of GRB prompt emission the PD is of $\sim (10\% - 40\%)$ \citep[e.g.][]{Deng16}, the detection of $\sim 10\%$ PD in GRB 090102 \citep{Steele09} and $\ge 28\%$ in GRB 120308A \citep{Mundell13} at the early stage of the RS dominant phase\footnote{In this paper, we do not simulate the PD and PA data of GRB 120308A  \citep{Mundell13}, which requires a more detailed simulation of a conical jet interacting with an ambient medium with the emission from both the RS and FS represented. Such a simulation is beyond the scope of this paper.} suggest that the PD is not reduced significantly in the RS region. Since turbulence is ubiquitous in a GRB environment \citep{Zhang&Yan11}, the preservation of such a moderately high degree of PD suggests a relatively strong $B$ field in the RS region. The low-$\sigma$ cases (e.g. $\sigma_4 < 10^{-3}$) may be ruled out. This conclusion is consistent with light curve modeling of GRB early afterglows that shows a relatively large magnetic microphysics parameter ratios between FS and RS \citep{Fan02,Kumar03,Zhang03,Mimica10,Gao15}.

Other effects not included in the current simulations include the possible RT instability in a conical jet geometry, the initial more complicated magnetic configuration directly adopted from our previous simulations of prompt emission \citep{Deng15}, and the 3D effect that in principle can lead to more tangled fields. We plan to investigate these effects in future more detailed studies. In any case, all of these parameters tend to further destroy the ordered magnetic field configurations and reduce the PD in the RS emission. The PD values obtained in this paper are conservative upper limits. The $\sigma$ constraints using the results obtained from this paper can be regarded as conservative lower limits.

\medskip
\acknowledgments
We are grateful to Christopher White, Xuening Bai, Gustavo Rocha da Silva, Zhaohuan Zhu, Jiming Shi, Yanfei Jiang and Chang-Goo Kim for the important technical support during the usage of the ``Athena++" MHD code, Haocheng Zhang for the helpful instruction regarding the ``3DPol" polarization calculation code, and Fan Guo, Haocheng Zhang, Xiangrong Fu, and Shangfei Liu for helpful discussions and suggestions.  An anonymous referee is greatly appreciated for very helpful comments. This work is supported by NASA through grants NNX15AK85G and NNX14AF85G, and by the LANL/LDRD program and Institutional Computing Programs at LANL and by DOE/Office of Fusion Energy Science through CMSO.

\end{CJK*}
\end{document}